# Formation of complex films with water-soluble CTAB molecules

S. Biswas [a, b], S. A. Hussain [a], S. Deb [a], R. K. Nath[b], D. Bhattacharjee [a]

Department of Physics [a] and Department of chemistry [b],
Tripura University, Suryamaninagar-799130, Tripura, INDIA

**Abstract:** This communication reports the formation of complex Langmuir monolayer at the air-water interface with the water-soluble $N-Cetyl\ N,N,N$-trimethyl ammonium bromide (CTAB) molecules when interacted with the stearic acid (SA) molecules. The reaction kinetics of the formation of the CTAB-SA complex was monitored by observing the surface pressure versus time graph. Multilayered LB films of this complex doped with congo red was successfully formed onto a quartz substrate. UV-Vis absorption and steady state fluorescence spectroscopic characteristics of this doped LB films confirms the successful incorporation of congo red molecules in to the CTAB-SA complex films.

* Corresponding author
E.mail: sa_h153@hotmail.com
Phone: +91-381-2375317
**Fax: +91-381-2374801**

**Introduction**:

The Langmuir-Blodgett (LB) technique of film deposition method is one of the most versatile techniques for fabrication of organized molecular assemblies in mono- and multilayer films, the architecture of which can be controlled by changing various LB parameters [1-4]. Due to these features, the LB technique is considered particularly suitable to handle and assemble different molecules with various chromophoric functions in a desired manner with the aim of fabricating molecular electronic devices as well as also to investigate various physico-chemical processes occurring at the mono- or multilayer. Typical LB compatible materials are amphiphilic molecules with long alkyl chain as a tail part and a hydrophilic head group. Well-defined mono- and multilayer LB films have also been prepared using lightly substituted molecule or molecules without any alkyl chain. Such nonamphiphilic molecules were incorporated into LB films when they were mixed with a long chain fatty acid (namely, stearic acid (SA) or arachidic acid (AA)) or an inert polymer matrix (namely, polymethyl methacrylate (PMMA)) [3-6]. The photophysical characteristics of such mixed LB films can be manipulated with ease, by changing various LB parameters.

In recent time, some water soluble materials were also observed to form well organized Langmuir monolayer at the air-water interface and suitably deposited onto solid substrates to form mono and multi layered LB films. However for this purpose an interaction with these water-soluble materials with another material is required and some LB compatible complex is formed in the process of interaction. Anionic and cationic surfactants are best examples of water soluble materials which sometimes form well organized Langmuir monolayer at the air–water interface when interacts with some insoluble materials. The architecture and molecular engineering of artificial supra molecular assemblies at surfaces or interfaces, in particular of complex monolayers with desired structure and physical properties have attracted great interest in recent years [7-8]. This complex monolayer consists of various components and the formation of this complex monolayer at the air-water interface depends on the interaction between the various components.

This communication reports a unique observation of the reaction kinetics between water-soluble $N-Cetyl\ N,N,N$-trimethyl ammonium bromide (CTAB) and stearic acid (SA) to form a stable complex monolayer at the air-water interface. The photophysical characteristics of the Langmuir-Blodgett (LB) films of this complex when doped with congo red (CR) have also been reported.

**Experimental**:

CTAB and Congo red (CR) were purchased from Loba chemie, Mumbai, India. Stearic acid (purity>99%) purchased from Sigma chemical Company were used as received. Chloroform used was of spectroscopic grade and its purity was checked by fluorescence spectroscopy before use. Langmuir-Blodgett film deposition instrument (Apex-2000C, India) was used to study the surface pressure vs. area per molecule isotherm characteristics and the reaction kinetics by studying the increase in surface pressure with time. Triple distilled deionised water was used as subphase and the temperature was maintained at 24º C. The SA solution was first spread on the water subphase by a micro syringe. After a delay of 15 minutes, to evaporate the solvent, the SA film at air-water interface was compressed slowly at a rate of $2\times10^{-3}$ nm$^2$mol$^{-1}$s$^{-1}$ to record the surface pressure versus area per molecule isotherm. When a desired surface pressure was achieved, CTAB solution was spread on the water of the LB trough from the backside of the barrier and the corresponding increase in surface pressure with time was recorded.

For deposition of LB films, we have first formed a stable Langmuir monolayer of SA and then the pre-mixed solution of CTAB and congo red (CR) was spread and sufficient time was allowed to interact with each other and finally the stable Langmuir monolayer at a desired surface pressure was transferred onto a quartz substrate following standard technique [1].

**Results and Discussion:**
**Formation of CTAB-SA complex monolayer at the air-water interface:**

For the observation of reaction kinetics the Langmuir-Blodgett (LB) instrument was used. 200 µl of dilute solution of SA (0.5 mg/ml in chloroform) was spread at the air-water interface of the LB trough and after sufficient time was allowed to evaporate the solvent, the barrier was compressed slowly to obtain the initial desired pressure. At that pressure the barrier was kept fixed and a dilute solution of CTAB in various amount was injected slowly into the water from the backside of the barrier. Since CTAB is water soluble, it was mixed with the water completely and reaction kinetics was started to form SA-CTAB complex.

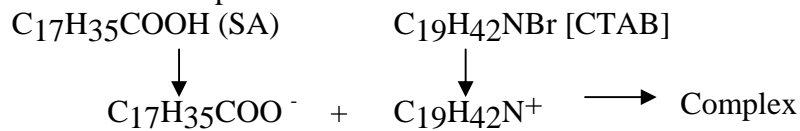

$$C_{17}H_{35}COOH \text{ (SA)} \qquad C_{19}H_{42}NBr \text{ [CTAB]}$$
$$\downarrow \qquad\qquad \downarrow$$
$$C_{17}H_{35}COO^- \;+\; C_{19}H_{42}N^+ \longrightarrow \text{Complex}$$

This complex of SA and CTAB was water insoluble. Area per molecule of this complex is greater than the stearic acid area per molecule. However as the barrier was kept fixed, area per molecule of the monolayer could not be increased, and as a consequence surface pressure began to rise and with increasing time the surface pressure showed a gradual rise which was an indication of reaction kinetics and formation of a stable complex monolayer at the air-water interface.

Figures 1a and 1b show the surface pressure versus time graph of this complex monolayer with various amount of CTAB spread on the water for two different initial starting surface pressure namely 0 mN/m and 15 mN/m respectively.

From these two figures it is observed that the surface pressure rises sharply with increasing the amount of CTAB. From figure 1a it is observed that from an initial surface pressure of 0 mN/m when the added amount of CTAB solution is very low, 100 µl, the surface pressure does not rise at all. With increasing the amount of solution the surface pressure started to rise gradually (graph 2 & 3) and when the added amount of CTAB solution is greater than 1000 µl, the surface pressure rises more than 40 mN/m. Also time taken becomes gradually smaller and when the added amount of CTAB solution is 2000 µl the surface pressure rises upto 40 mN/m from an initial value of 0 mN/m with in a time span of only 2.75 hours (graph 8).

In figure 1b the initial surface pressure of pure stearic acid monolayer is 15 mN/m. When the CTAB solution was added the surface pressure gradually rises and when the added amount of CTAB solution is 2000 μl, the surface pressure rises to 40 mN/m from an initial 15 mN/m with in a time span of 2.12 hours. This clearly indicates that some complex of SA and CTAB molecules are formed at the air-water interface. This complex molecule is water insoluble and forms a complex monolayer. The area per molecule of this complex monolayer is greater than the stearic acid area per molecule. However since the barrier is kept fixed at a particular position, surface pressure gradually increases. This increase in surface pressure with time is actually a direct evidence of reaction kinetics between SA and CTAB molecules. Figure 2 shows a schematic representation of the CTAB-SA complex Langmuir monolayer at the air-water interface.

**Deposition of multilayered LB films from the Langmuir monolayer of CTAB-SA complex doped with Congo red (CR):**

CTAB, SA as well as CTAB-SA complex don't have any absorption or fluorescence characteristics. Therefore to get an idea about the formation of stable LB films, Congo red (CR), a highly fluorescent dye material was doped in the Langmuir monolayer of CTAB-SA complex at the air-water interface. Congo red, an anionic molecule interacts with the cationic part of the CTAB molecules and thus forms a complex.

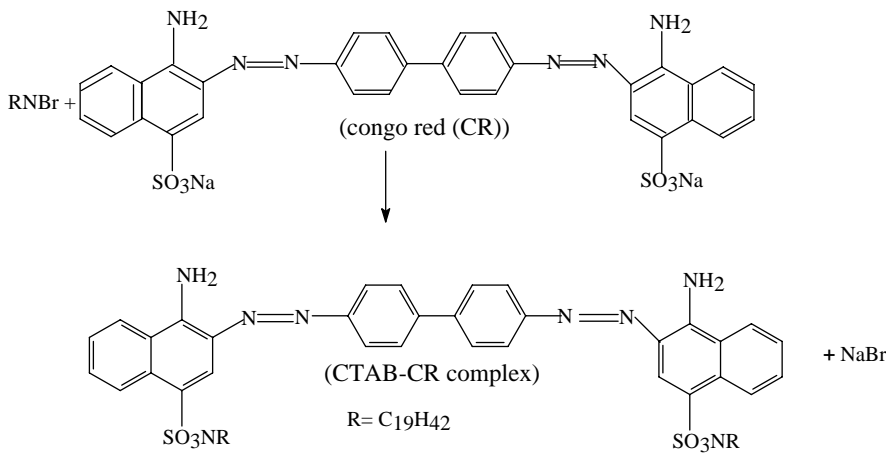

The interaction between CTAB and Congo red is also evident from the absorption and the fluorescence spectroscopic study (fig. 4a and 4b) in the mixed solution phase. However, no interaction occurred between Congo red and SA molecules. When a mixed solution of CTAB and CR having excess amount of CTAB (CTAB: CR=80:20) is added to the water surface of the Langmuir trough, where a previously formed SA monolayer exists, then the excess amount of CTAB interacts with the SA molecules of the Langmuir monolayer and forms a CTAB-SA complex monolayer. Within these complex molecules, the CTAB-CR complex molecules tend to get sandwiched. Thus the ultimate Langmuir monolayer becomes a composite form of a complex of CTAB-SA: CTAB-CR: CTAB-SA. Ultimately the multilayered Langmuir Blodgett films formed from this complex monolayer also consists of SA, CTAB and Congo red molecules. Figure 3 shows a schematic representation of the CTAB-SA complex Langmuir monolayer doped with congored at the air-water interface.

**Spectroscopic Characterizations of complex LB films:**

Figures 4a and 4b show the UV-Vis absorption and fluorescence spectra of LB films of CTAB-SA doped with congo red (CR) along with CR, CTAB-CR solution and CR microcrystal spectra for comparison.

CR solution absorption spectrum gives a longer wavelength intense band at 499 nm alongwith a low intense high-energy band at 340 nm. Mixed solution absorption spectrum of CTAB-CR gives a blue shifted longer wavelength intense band with peak at 473 nm as well as also a high-energy blue shifted band with peak at 329 nm. This clearly suggests an interaction with anionic CR and cationic CTAB. CR microcrystal absorption spectrum gives a broad low intense longer wavelength band with peak at around 513 nm. The absorption spectrum of LB film doped with CR, gives prominent peak of CR. This clearly suggests that CR is successfully doped in the LB films of CTAB-SA complex.

Fluorescence spectra (figure 4b) also show that the LB films spectra is almost similar to CR microcrystal spectrum. This clearly suggests the formation of SA-CTAB-CR complex LB films.

**Conclusion**:

In conclusion our result shows that water-soluble CTAB molecule can be successfully organized into Langmuir monolayer when interacted with SA molecules. Thus forms a complex Langmuir monolayer at the air-water interface. The reaction kinetics of the formation of the CTAB-SA complex was monitored by observing the surface pressure versus time graph. Multilayered LB films of this complex doped with CR was successfully formed. UV-Vis absorption and steady state fluorescence spectroscopic characteristics of this doped LB films confirm the successful incorporation of CR molecules in the CTAB-SA complex films.

**Acknowledgement**
The authors are grateful to DST, Government of India for providing financial assistance through SERC-DST Project No. SR/FTP/PS-05/2001.

**References:**

1. A. Ulman, An introduction to Ultrathin Organic Films: From Langmuir–Blodgett Films to Self Assemblies, Academic Press, New York, 1991.
2. M.C. Petty, Langmuir–Blodgett Films: An Introduction, Cambridge University Press, 1996.
3. S. Deb, S. Biswas, S. A. Hussain, D. Bhattacharjee, Chem. Phys. Lett. 405 (2005) 323
4. S. Acharya, D. Bhattacharjee, J. Sarkar, G. B. Talapatra, Chem. Phys. Lett. 393 (2004) 1
5. S.Acharya, D.Bhattacharjee, G.B.Talapatra, Chem. Phys. Lett. 372, (2003) 97
6. A. G. Vitnkhnovsky, M. I. Sluch, J. G. Warren, M. C. Petty, Chem. Phys. Lett. 173 (1990) 425
7. M. Ferreira, C. J. L. Constantino, A. Rlul Jr., K. Wohnrath, R. F. Aroca, J. A. Glacometti, O. N. Oliveira Jr., I. H. C. Mattogo, Polymer 44 (2003) 4206
8. M. Ferreira, R. L. Dinelli, K. Wohnrath, A. A. Batista, O. N. Oliveira Jr.Thin Solid Films 446 (2004) 301

**Figure caption:**

Figure 1(a): Surface pressure versus time graph of the CTAB-SA complex monolayer with various amount of CTAB spread on the water for initial starting surface pressure of 0 mN/m. 1=100, 2= 250, 3= 500, 4 = 750, 5 = 1000, 6 = 1250, 7 = 1500, 8 = 2000 µl of CTAB.

Figure 1(a): Surface pressure versus time graph of the CTAB-SA complex monolayer with various amount of CTAB spread on the water for initial starting surface pressure of 15 mN/m. 1= 250, 2= 500, 3 = 750, 4 = 1000, 5 = 1250, 6 = 1500, 7 = 2000 µl of CTAB.

Figure 2: (a) Schematic representation of Langmuir monolayer of SA at the air-water interface. (b) Schematic representation Langmuir monolayer of CTAB-SA complex at the air-water interface.

Figure 3: Schematic representation of Langmuir monolayer of CTAB-SA complex at the air-water interface doped with CR.

Figure 4(a): UV-Vis absorption spectra of LB films of CTAB-SA doped with congo red (CR) along with CR, CTAB-CR solution and CR microcrystal spectrum.

Figure 4(b): Fluorescence spectra of LB films of CTAB-SA doped with congo red (CR) along with CR, CTAB-CR solution and CR microcrystal spectrum.

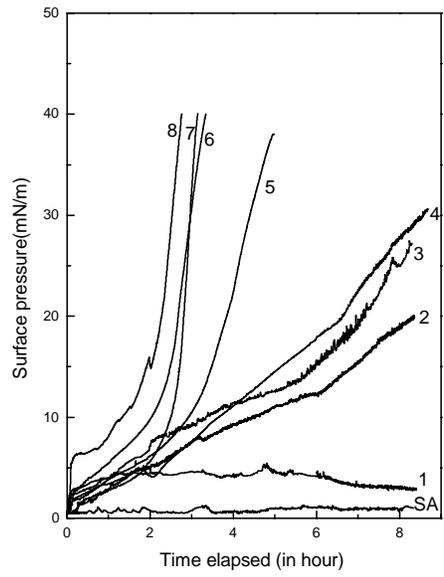

Figure 1(a). S. Biswas et. al.

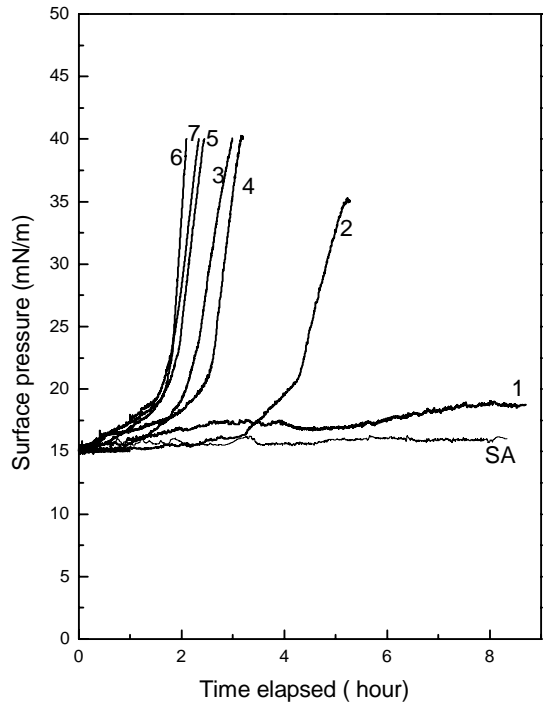

Figure 1(b) S. Biswas et. al.

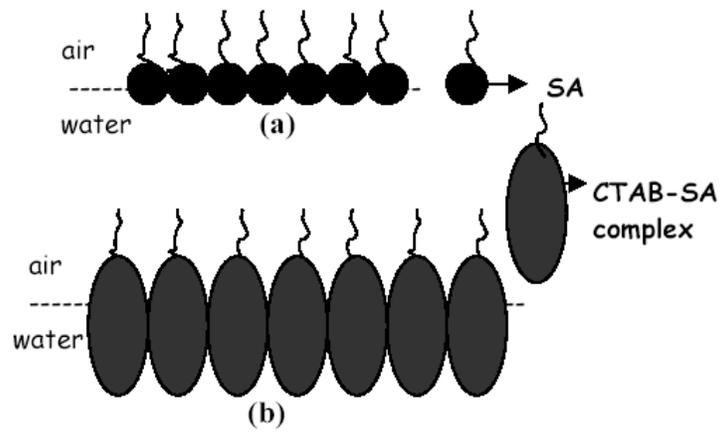

Figure 2. S. Biswas et. al.

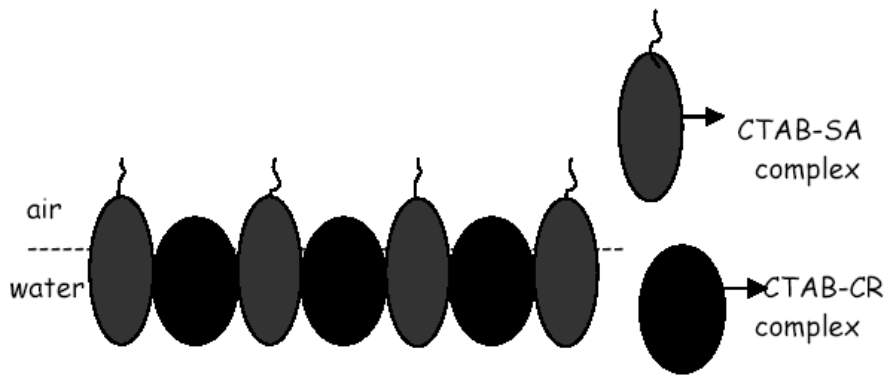

Figure 3 S. Biswas et. al.

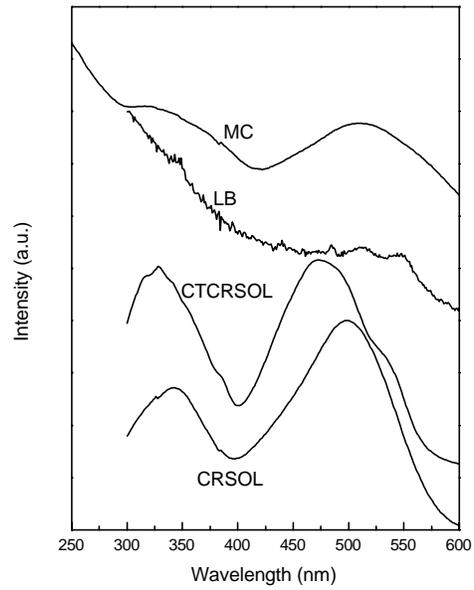

Figure 4(a) S. Biswas et. al.

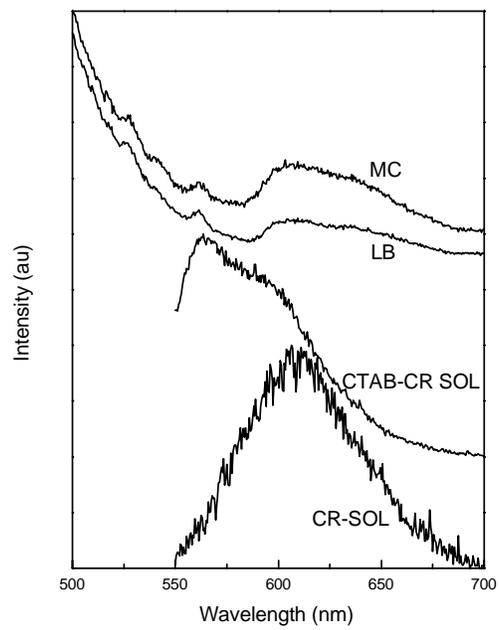

Figure 4(b) S. Biswas et. al.